# Steganography Security: Principle and Practice

## Yan Ke*, Jia Liu, Min-qing Zhang, Ting-ting Su and Xiao-yuan Yang

Key Laboratory of Network and Information Security under the Chinese People Armed Police Force (PAP), College of Cryptography Engineering in Engineering University of PAP, Xi'an, 710086, China

Corresponding author: Yan Ke (e-mail: 15114873390@ 163.com).

This work was supported in part by the National Key Research & Development Program of China under Grant No. 2017YFB0802000, and Natural Science Foundation of China under Grant 61379152 and Grant 61872384.

**ABSTRACT** This paper focuses on several theoretical issues and principles in steganography security, and defines four security levels by analyzing the corresponding algorithm instances. In the theoretical analysis, we discuss the differences between steganography security and watermarking security. The two necessary conditions for the steganography security are obtained. Under the current technology situation, we then analyze the indistinguishability of the cover and stego-cover, and consider that the steganography security should rely on the key secrecy with algorithms open. By specifying the role of key in steganography, the necessary conditions for a secure steganography algorithm in theory are formally presented. When analyzing the security instances, we have classified the steganalysis attacks according to their variable access to the steganography system, and then defined the four security levels. The higher level security one has, the higher level attacks one can resist. We have also presented algorithm instances based on current technical conditions, and analyzed their data hiding process, security level, and practice requirements.

## I. INTRODUCTION

The modern secret communication technology originated from the military demands of the secret communication during World War II, and the realization of secret communication mainly consists of two major technologies: *cryptography* and *steganography*. From the viewpoint of security requirements, the significance of cryptography is to maintain the secrecy of the communication contents while that of steganography lies in the undetectability of the existence of the secret communication. From the viewpoint of the attacker or the analyzer, the purpose of the cryptanalysis is to finally obtain the decryption key or plaintext sequences while steganalysis is to detect the presence of the communication which does not require any communication content. Therefore, for the two security systems, security in steganography is more fragile than cryptography, i.e., once the existence of the communication is exposed or detected, security could be compromised. In this paper, steganography security is our main attention. The security levels of steganography are classified by introducing different models of attacks. And practical instances at different levels are present theoretically and practically based on existing data hiding technologies.

Data hiding technology has evolved into two branches, watermarking and steganography. Their prototypes were first defined with modern terminology in Simmons' founding work on subliminal channels and the prisoners' problems [1]. We next analyze the differences existing in their security requirements, so as to accurately present the requirements of steganography security.

In prisoner's problems [1], Alice and Bob are in jail and they want to devise an escape plan by exchanging hidden messages in innocent-looking covers (e.g., natural images). These covers are conveyed to one another by a common warden, Eve, who can eavesdrops all covers and can choose to interrupt the communication if they appear to be a stego-cover. In this model, the secret communication consists of three basic elements: 1 *secret message*, 2 *cover*, and 3 *open channel*. The open channel has been assumed to be the worst case for steganography security, since the existence of the channel is implemented by Eve. Eve has access to all the contents that Alice and Bob have exchanged, and she can also interrupt the channel if she finds the exchange content is not normal. Then the steganography security can be considered as breach. However, the exposure of the secret communication is not a sufficient condition to breach the watermarking security, since watermarking is mainly used for the authentication and anti-counterfeit of the copyrights of the covers in practical applications. The presence of the watermark is directly exposed in the visible watermarking, or some copyright owners choose to disclose the watermark labels previously for future extraction and certification.

Under the watermarking mode, Eve's task is not only to detect the presence of the secret communication, but also to realize the destruction, tampering, and forging on the watermark. She needs to complete the goal of an active attacker, which is consistent with the definition by Kalker [2]: "Watermark security refers to the inability by unauthorised users to have access to the raw watermarking channel." Under the steganography mode, Eve only needs to complete the goal of a passive attacker while the actual attack could be active behaviors. It means that Eve can perform active accumulation from the system aiming to detect the presence of the secret communication. Therefore, the steganography system gives Eve more active privileges to complete a passive attacking goal. From the perspective of the data hider (Alice or Bob), it calls for some higher security requirements to resist Eve in the steganography system.

In conclusion, we can give the definition of the *steganography security* (*stego-security* for short):

*Definition* 1, it can be considered as *stego-security* if it could ensure (passive or active) Eve cannot complete the following two attack goals:

1. *To detect that the stego-cover is unnatural;*
2. *To extract the contents of the hidden message.*

The above two attacks can be interpreted that the existence of the secret communication is exposed from the perspective of the cover or the secret message, respectively. The occurrence of each exposure means the security breach. In Section V, we will continue to discuss them.

It should be mentioned that J. Fridrich classified Eve's behaviors into *passive behavior*, *active behavior* and *malicious behavior* [3]. The malicious behavior refers to an attack that Eve might destroy or tamper with the channel even when she is not aware of the existence of the secret communication. The main thing to consider then is robustness, not security, since such attack is not a threat to the security, even though the current channel is being vandalized. In the reality of multimedia and the network environment, the data hider can reselect another open channel, and it doesn't consider the situation of no available channel. And also, the robustness is not the focus of this paper.

The rest of this paper is organized as follows. The following section discusses two stego-security principles of current technology. Section III introduces the security requirments of key in steganography. Section IV reviews the different theoretical models of data hiding, and then Section V describes the four levels of steganalysis attacks. In Section VI, the stego-security instances are discussed theoretically. Finally, Section VII summarizes the paper and discusses future investigations.

## II. SECURITY PRINCIPLES OF STEGANOGRAPHY

The steganography cover is mainly digital images. Therefore, we analyze the following two security principles mainly based on image steganography in this section.

### A. NOTATION

In this paper, Cover, denoted as $C$, is used to carry secret messages and conceal the existence of the secret communication, such as images, videos, texts, etc.

Stego-cover, denoted as $C'$, is the cover that has been embedded with the secret message.

Secret message, denoted as $m$, is the message to be embedded into the cover for communication.

### B. INDISTINGUISHABILITY

The indistinguishability of cover and stego-cover is an important guarantee of stego-security. For the attackers, it is a variety of covers that they can directly obtain in the open channels. Current steganography methods also focus on the indistinguishability. Due to the complexity of data hiding process and the texture of the cover image, it is unlikely that the message will be directly extracted by the attacker. Therefore, without considering the extraction of the secret message, the indistinguishability can be directly equivalent to the stego-security [4]-[7]. To discuss this issue, we give an explanation of the indistinguishability of cover and stego-cover.

Firstly, the concept of an abstract distance $D$ ($D \geq 0$) in this paper is given, which quantifies the differentiability between different samples.

$D(P(C), P(C'))=0$ indicates that the covers $C$ is indistinguishable from the stego-covers $C'$, which means not only the *imperceptibility* in the human visual system, but also the *undetectability* in the sense of statistical analysis.

*Definition* 2, $C$ and $C'$ are indistinguishable iff (if and only if)
$$D(P(C), P(C'))=0 \qquad (1)$$

*Definition* 3, $C$ and $C'$ are distinguishable iff
$$D(P(C), P(C'))>0 \qquad (2)$$

In our analysis of the security, the standard of indistinguishability is $D=0$ theoretically, but in practical applications, it is usually not directly evaluated by $D=0$. Here, $(t, \varepsilon)$-*distinguishable* is defined:

*Definition* 4, $C$ and $C'$ are $(t, \varepsilon)$-distinguishable iff
$$D(P(C), P(C'))>\varepsilon \qquad (3)$$
in running time $t$, in which $\varepsilon$ is an arbitrary small value.

There are many specific mathematical tools to indicate the distance $D$. For the human visual system, there are mature measures used in the audiovisual quality of individual experience. ITU-T has defined several ways of subjective evaluation criteria, such as ITU-T Rec. P.910, this recommendation is intended to define non-interactive subjective assessment methods for evaluating the quality of digital images. Mean Opinion Score (MOS) is the most representative subjective evaluation method for image quality, which can judge the image quality by normalizing the evaluations of the observers [8]. For the statistical analysis, Cachin used kullback-leibler divergence distance (KL divergence distance, $D_{KL}$) [4] to describe the distribution deviation. The formalized expression of $D_{KL}$ in steganography system was given by using the information entropy theory [3], [5]. In [6], F. Cayre first reviewed several forms of watermark attacks, and then introduced conditional probability to analyze the watermarking security level in watermarked only attack (WOA), and gave a formalized expression of the highest watermarking security which was referred to as stego-security:

$$P(C'/K)) = P(C) \qquad (4)$$

where $P(C'/K)$ denotes the model of $C'$ by data hiding key $K$. Then it is derived that Eq. (4) is equivalent to Eq. (1) by KL divergence distance. In [7], some security concepts were inspired by a cryptographic approach. It assumed that there was an attack to distinguish the different distribution models in polynomial time, the distinguishing probability of the attack was used to quantify the deviation of distributions. In the field of generative adversarial networks (GANs), the discriminator for measuring deviations between samples includes: Jensen-Shannon distance (JS distance, $D_{JS}$) and Wessertein distance (W distance, $D_W$). W distance has advantage over the JS distance in optimizing the distribution of the output from the generators [9]. Since the imperceptibility in the human visual system mainly depends on the subjective evaluation, we do not take any more analysis. The main consideration of indistinguishability concentrates on the undetectability under the statistical analysis.

Following the same distribution under existing criteria (e.g. KL divergence distance, etc.) is not equivalent to the indistinguishability of cover and stego-cover (i.e., Eq. (5)). We can analysis that by following three reasons: 1) The results of the distribution fitting under the existing criteria can be compromised due to the quantity and effectiveness of the sampling data; 2) The *imperceptibility* and *undetectability* cannot always be assured simultaneously in state of the art; 3) In view of the large data amount of the natural image (even if a single image contains millions of pixels), assuming that there is a distribution model of the natural image, the model has such a high dimension and an extreme difficulty to be understood that it is far from being realized by the current computational power and storage capacity in reality. However so far, KL distance and its variations have been the current best criteria of *undetectability* [3], [5].

$$D(P(C), P(C')) = 0 \Rightarrow \begin{cases} D_{KL}(P(C), P(C')) = 0 \\ D_{JS}(P(C), P(C')) = 0 \\ D_W(P(C), P(C')) = 0 \\ ... \end{cases} \qquad (5)$$

And according to the equivalence of the inverse negative proposition, it can be deduced:

$$\left. \begin{array}{l} D_{KL}(P(C), P(C')) \neq 0 \\ \text{or } D_{JS}(P(C), P(C')) \neq 0 \\ \text{or } D_W(P(C), P(C')) \neq 0 \\ ... \end{array} \right\} \Rightarrow D(P(C), P(C')) \neq 0 \qquad (6)$$

Eq. (6) indicates that the if there is a deviation between cover model and stego-cover model under the existing distance criteria, there will be sufficient reasons to theoretically prove that the steganographic method cannot ensure the indistinguishability.

In [3], J. Fridrich classified the steganographic methods into three main categories: cover-selection, cover-synthesis, and cover-modification. The cover-modification based steganography is the main body of steganographic techniques. The other two categories are concluded as *non-modified steganography* in this paper. According to information theory, there are artificially irreversible changes added into the cover in cover-modification based steganography, thus resulting in a distinct deviation when fitting the distributions of cover model and stego-cover model under KL distance [3].

Even so, however, it does not mean that the modified cover must be insecure, because in practice, the attacker's computational ability is also limited. Moreover, if the probability of a successful analysis decreases exponentially with the increase of the cover size or the key length in steganography, it is possible to construct a data hiding method that meets certain complexity requirements and security needs in practical applications, namely, the calculating complexity of the data hiding method can be increased until the attacker cannot finish the analysis within the polynomial time [10][11]. On the other hand, since it is extremely difficult to ensure the indistinguishability with the natural image after modification, a steganographic framework, which allows the finite embedding distortion, is proposed [12]. Thereafter, it continues to evolve to the steganography by minimizing additive distortion using Syndrome-Trellis codes (STC) [13], such as HUGO [14], WOW [15], MVGG [16], and UED [17].

However, with deep learning technology introduced into steganalysis, its efficient learning ability for high-dimensional and high-complexity models has continuously reduced the availability of the algorithms that are designed to meet complexity requirements or finite distortion [18][19]. The modification that occurs in the cover without mastering the cover model has been becoming a hidden threat to stego-security.

In conclusion, the cover-modification based steganographic methods cannot ensure the indistinguishability in theory, but can resist certain analysis attacks and meet some security requirements in practice. However, due to the theoretical natural defects in security, with the development of analytical tools, the analysis risk of such methods will gradually increase. In view of the future development of the secure steganography, this paper aims to consider methods that are capable of ensuring the indistinguishability of cover and stego-cover theoretically and practically. Therefore, besides optimizing the security of the cover-modification based methods, we also pay attention to the following three data hiding methods:

1. Steganography based on *non-modified steganographic methods*, whose characteristic is that no modification

occurs in the cover after embedding, can effectively guarantee the indistinguishability, such as the coverless information hiding, and generative steganography based on GANs.

2. Due to the difficulty of building a perfect model of natural images, the options of the cover can be further broaden, especially the ones with simple unified distribution models, such as reversible data hiding in encrypted domain (RDH-ED) in the cloud environment. In RDH-ED, the cover is the encrypted data that just follows the random uniform distribution. The indistinguishability essence is to ensure the stego-ciphertext to follow the random uniform distribution.

3. The cover of immeasurable modeling in emerging technologies can be selected for steganography, such as steganography based on the current quantum technology [20][21].

However, considering the current technique conditions, steganography based on the above methods 2 and 3 is controversial, because their covers have not been widespread (The encryption technique in cloud has not been standardized, and the quantum technique is still at the experimental stage). We enumerate those methods only in view of theoretical analysis.

### C. ALGORITHM-SECRECY OR KEY-SECRECY BASED SECURITY

After ensuring the indistinguishability, the steganographic methods need to consider another issue: whether the security should be based on the *secrecy of algorithms* or the *secrecy of the key* with the algorithm open. The basis of modern system security usually develops from algorithm secrecy to key secrecy with algorithm open, which is mainly the choice of the technology development in practice. Algorithm-secrecy based security has several drawbacks. We can discuss it from the following three aspects:

1. The cost of keeping secrecy. The cost of algorithm-secrecy based security is relatively high, which includes: the algorithms, system software, system hardware, operating personnel, and etc.

2. The cost of updating system. Limited algorithms can be supported by one electronic product, so the system needs updating frequently. To achieve the same security, algorithm- secrecy based security system needs to update algorithm, hardware, and personnel operator skills, etc., while key-secrecy based security system only needs to update another key.

3. The standardization of security system. Algorithm open is the precondition for qualitative analysis on system security, as well as the practical popularization. The security cannot be proved through open mathematical analysis, if the details of an algorithm are kept confidential. The method with indeterminate security cannot be applied directly to the standardization or industry popularization.

In conclusion, the practical security system should depend on the security under the condition of algorithm open. At present, open algorithm is a huge challenge, because it has to resist the specific steganalysis. Currently, few methods support that. Even STC steganography [13], which is generally recognized for its current performance, has a poor effect in resisting the specific steganalysis [22].

Current situations and technical issues in the above two aspects seem to predict the "two-step" development of steganography security: The first step is to ensure the indistinguishability after embedding; second, it should be based on key secrecy with algorithm open. Then, the steganalysis method would ultimately come down to the brute force attack to extract the embedded data directly. At this time, stego-security only depends on the size of the key space for message extraction. The introduction of the concept of key and key space, as well as the implementation in steganography can refer to one important principle of current cryptosystem, Kerckhoffs' principle [23]. In the next section, we mainly analyze the key in steganography.

### III. KEY IN STEGANOGRAPHY

Key is an important part of secret communication system, whose function is not only to extract the message. In cryptography, the key determines the visibility of the content in secret communication. Likewise, in steganography, the key should determine the detectability of the existence of the secret communication. The significance of this paper emphasizes that the function of key should formally embody in stego-security.

### A. KERCKHOFFS' PRINCIPLE IN STEGANOGRAPHY

In stego-security, it is the data hiding method that ensures the indistinguishability, while the secrecy of the hidden message content with the algorithm open relies on the usage principle and the secrecy of the key by Kerckhoffs' Principle. In 2002, Furon, *et al.*, have introduced Kerckhoffs' Principle into the data hiding field [24]. In [25], it classified the watermarking attacks according to Kerckhoffs' principle. On the basis of [25], F. Cayre defined four levels of watermarking security under WOA [6], of which the highest watermarking security was denoted as steganography security. But steganography security was not discussed any further. In [26], Kerckhoffs' Principle of the steganographic techniques was proposed. The framework in [13] also considered that in the case of algorithm open. Above all, we emphasized that the usage of the key in steganography scheme should be based on Kerckhoffs' Principle, i.e., the key-secrecy based stego-security only relies on the secrecy of the key, and any information except the key about the steganography system can be exposed to the attacker.

### B. KEY IN STEGANOGRAPHY

In Kerckhoffs based steganography system, the key is not the *procedural element* that reflects the function or purpose of the system, but the *security element* that relates to the complexity of the system. It means that certain process information of the specific algorithm (such as image block size, LSB or 2LSB as the embedding layer, etc.) should not be used as the key, though the secret message would not be extracted without such information, which in fact belongs to algorithm-secrecy based security.

Compared with procedural elements of the system, the key should have the biggest information entropy, i.e., the numerical uncertainty, so as to meet the requirement of a large key space. At the same data length, to maximize the information entropy, the data should follow the random uniform distribution instead of any regular one. Therefore, the key is always derived primarily or entirely from the random number generator. On the premise of the indistinguishability, the key-secrecy based stego-security is only determined by the size of the key space provided by the steganography algorithm. The key space can be quantified by the key length and computational complexity theory. The larger key space an algorithm can provide, the greater security it can provide (Note that: if the embedded data is encrypted with a cipher, the encryption key space (if any) provided by the cipher cannot be regarded as the steganography key space). We will continue to analyze the relationship between the key space and stego-security in our future work.

The necessary conditions of the key-secrecy based stego-security can be formally obtained:

Data hiding:
$$F_{stego}(C, m) \xrightarrow{k} C' \tag{7}$$

Data extraction:
$$F_{extract}(C', k) \rightarrow m, F_{extract}(C', .) \rightarrow m', \Pr(m = m') \leq 50\% \tag{8}$$

Security requirements:

1. $D(P(C), P(C')) = 0$.

2. $k \in \mathcal{K}$, $\mathcal{K} \subseteq \mathbf{R}$, $|\mathcal{K}|$ meets the certain requirement of computational complexity, where $k$ denotes the key, $\mathcal{K}$ represents the collection of all possible keys, $|\mathcal{K}|$ is the size of the key space, $\mathbf{R}$ denotes random number set, and $m'$ denotes the message obtained from the stego-cover without $k$.

The key discussed in this paper is basically in the symmetric system. For a public key system, it would require an additional condition that the embedding key can only be unidirectionally derived from the extraction key. The embedding key can be publicly released and the extraction key shall be kept secrecy [27].

## IV. CURRENT STEGO-SECURITY MODELS

Through the above introduction and analysis, we review the theoretical models in this part. Since Simmons' prisoners' problem was proposed in 1984, data hiding models have drawn much research attention. Existing data hiding theory models are concentrating more on watermarking security, in which stego-security just represents a securest state that watermark security could achieve in theory. Early in 2001, Kalker [2] proposed a definition of watermarking security. Then P. Comesana [28] in 2005 proposed an information-theoretic framework to study watermarking security and robustness, and used it to analyze the Spread-Spectrum (SS) data hiding method. In [25], the security was classified by introducing four attacking modes: known-message attack (KMA), known-original attack (KOA), watermarking-only attack (WOA), and constant-message attack (CMA). F. Cayre further defined the four levels of security under WOA [6]. In [29], Wang, et al. reviewed watermarking security levels and SS embedding methods under WOA. All above models have exemplified the security requirements of watermarking techniques and the interactions between the data hider (Alice, Bob) and the attacker (Eve). However, the behaviors and the security target of the interactions under watermarking mode are mostly inapplicable under steganography mode (as described in Section I). Stego-security was only considered as the highest security level in the watermarking models on extreme hypothesis [6][29]. The guiding significance of these models to steganography stays at the conceptual level with deficiency of certain interactive arguments and practical criteria to theoretically demonstrate stego-security.

On the other hand, there has also been research work concentrating on resisting the passive behaviors of Eve and evaluating the security using an information-theoretic model in [4]. Its security goal was consistent with stego-security. However, it did not take Eve's active behaviors into consideration and the setup of the attacks might be one-sided. Ker in [30] extended the setup by introducing the concept of batch steganography and pooled analysis in which covers could be accumulation to improve Eve's knowledge. The setup of the attacks in this paper (Section V) follows Ker's line to analysis the different accesses of Eve to the stego-system.

Researches on the indistinguishability of the natural cover and stego-cover proposed several theoretical criterions, which has been reviewed in Section II. B. In Table 1, we enumerate six representative data hiding models and the proposed one, and present their modes of the prisoners' problem, the attacking goals, the modes of the attacking methods, the theories of their security criterions, and whether the algorithms can be open (i.e. key secrecy based security).

In this paper, we attempt to further analysis the stego-security including the specific principles and the practical implementation techniques. As for our attacking setup, Eve only needs to complete a passive attacking goal by using active or passive attacking methods. If a steganography method can simultaneously ensure the indistinguishability and provide a large enough key space, the method can be considered stego-secure theoretically. But state of the art is far from that. From a practical point of view, we impose some certain restrictions on the reality conditions, and classify stego-security into different levels in the following sections.

**TABLE. 1. DATA HIDING MODELS.**

| Literature | Mode | Attacking goal | Attacking mode | Security criterions | Algorithm open or not |
|---|---|---|---|---|---|
| [6] | watermarking | active | passive | KL divergence distance | open |
| [4] | steganography | passive | passive | Conditional entropy | unknown |
| [30] | steganography | passive | active | Hypothesis testing on steganalysis probability | unknown |
| [10] | steganography | passive | passive | complexity on computational indistinguishability | open |
| [7] | steganography | passive | unknown | steganalysis probability | open |
| [14] | steganography | unknown | unknown | Minimized finite embedding distortion | unknown |
| *The proposed* | *steganography* | *passive* | *active* | *Definition 1 in Section I and the principles in Section II* | *open* |

## V. STEGANALYSIS ATTACKS

Before introducing the stego-security classification, we discuss the different levels of steganalysis attacks. According to game theory, there exists the unique equilibrium between two rational sides with certain prior knowledge and the fixed rules in game. The equilibrium state means that both sides have achieved their best rational use of prior knowledge and maximized their lowest benefits. In the game of steganography and steganalysis, the goal and the rules are fixed, so the maximum possible benefit of the attack is only associated with the prior knowledge of the attacker [31] [32] (related theoretical derivation we will discuss in detail in the future work). Therefore, we classify the attacks based on the prior knowledge about the steganography system that the attacker is assumed to have. Intuitively, the more information about the attacking target the attacker can obtain, the easier it is to achieve the attack goal, and we call it *the higher level attack*. The higher level attack a steganography algorithm can resist, the higher level security the algorithm possesses.

As introduced in Section I, there are two criteria of completing the attack goal: one is to detect the stego-cover unnatural. The other is to extract messages from the stego-cover. The first focuses on the cover, and is currently the main starting point of steganalysis based on statistical modeling and feature analysis. The second focuses on the secret message, whose starting point is not to distinguish the cover, but to attempt to directly extract meaningful or sensitive information from the cover under the conditions of the available access and the open knowledge, then it may directly determine the presence of steganography.

This paper draws on the lessons of the implementation of cryptanalysis and divides the steganalysis attack into two phases: *the learning phase* and *the challenge phase*. In the learning phase, the attacker learns to analysis the steganography method based on current prior knowledge. In the challenge phase, the attacker would randomly obtain a natural cover or a new stego-cover when the learning phase is completed. The challenge would be considered successful, if he can distinguish the stego-cover or extract the message to determine the presence of steganography with a probability bigger than 50%. The challenge would be considered failed, if the probability is no bigger than 50%.

An attack that distinguishes the stego-cover from the original ones can be expressed in the following form:

Based on *Definition* 2,
$$\Pr[\ D(P(\boldsymbol{C}), P(\boldsymbol{C'})) \neq 0] > 50\% \quad (9)$$

Based on *Definition* 3,
$$\Pr[\ D(P(\boldsymbol{C}), P(\boldsymbol{C'})) > \varepsilon] > 50\% \quad (10)$$

An attack that recovers the secret messages from the stego-cover can be expressed in the following form:

$$\Pr(\boldsymbol{m} = \boldsymbol{m}') > 50\% \quad (11)$$

Based on the prior knowledge that the attacker has in the learning phase, the following four levels of attacks are defined, and the level increases in turn.

### A.  STEGO-COVER ONLY ATTACK (SCOA)

SCOA is the primary level attack model where the steganalysis attacker is assumed to have access only to a set of stego-covers. While the attacker has no channel providing access to the corresponding original cover prior to data hiding, in all practical SCOA, the attacker still has some knowledge of the existing natural covers.

In the learning phase, the attacker can use the statistical analysis method to model the distribution of the natural covers. The purpose is to analyze the covers in the open channel and then determine the presence of steganography. Corresponding to the case in reality, the attacker may monitor all public channels.

After completing the SCOA learning phase, the attacker enters the challenge phase.

### B. KNOWN COVER ATTACK (KCA)

KCA is an attack model for steganalysis where, in addition to the prior knowledge under SCOA, the attacker also has several pairs of the original cover and its corresponding stego version. The number of pairs is finite within the polynomial complexity.

In the learning phase, the attacker can not only carry out the learning in SCOA, but also learn from the pairs of the original cover and their stego version. Corresponding to the cases in reality, in addition to monitoring all open channels, the attacker can also steal original samples by hacking into the steganography system, or send spies to steal the original samples from the inside.

After completing the KCA learning phase, the attacker enters the challenge phase.

### C. CHOSEN COVER ATTACK (CCA)

In addition to the prior knowledge under the KCA, the attacker can also invoke several times of embedding or extraction process of the current steganography system. The number of the invoking operation is finite within the polynomial complexity. (It should be noted that the embedding and extraction algorithm is already open, and the invoking operation is to invoke several times of the key used for the present secret communication while the key itself remains secret.)

In the learning phase, in addition to the KCA learning phase, the attacker can invoke the embedding or extraction process to learn the changes of the cover. The cover used for extraction can be a stego-cover or a specific forged cover by the attacker. Corresponding to the cases in reality, the attacker can remotely control the steganography system to implement data embedding or extraction for several times, or instigate the user of the system to return several operation results.

After completing the CCA learning phase, the attacker enters the challenge phase.

### D. ADAPTIVE CHOSEN COVER ATTACK (ACCA)

After the CCA challenge phase, if the attack fails, the attacker can continue to restart the CCA learning phase on the targeted system, and can repeat the learning and the challenge phases several times until the attack succeeds.

In fact, SCOA can be regarded as the passive attack while the other higher level attacks are the active attacks.

The above classification is mainly based on the attackers' accesses to the cover rather than their different usage rights of the secret message, because it is the covers that the attacker can directly obtain in the channel, and the steganography technology mainly relies on the change of the cover to deliver additional information. For the secret message, data hider does not care about its content, and the content does not provide attackers with much reference for future attacks because the message can be encoded or encrypted before embedding. If it is random-like, the influence on the cover may be indistinguishable.

## VI. STEGO-SECURITY CLASSIFICATION AND INSTANCES ANALYSIS

This section gives instances for explaining the stego-security classification. An algorithm that can resist a high-level attack is necessarily resistant to the lower-level attacks.

### A. STEGO-SECURITY AGAINST SCOA

#### 1) COVERLESS STEGANOGRAPHY

The main characteristic of coverless steganography is that there is no pixel modification on the image exchanged in the channel. This section formalizes an abstract coverless steganography method based on the method in [33] as an instance to analyze its stego-security and practical requirements.

*a. Algorithm*

1. Sender and receiver share an image library $X$ containing $T$ natural images, $X$: $X=\{c_1, c_2, c_3,\ldots, c_T,\}$;

2. Select $N$ ($N<T$) images randomly from $X$, and arrange them into a permutation, $\{c_1', c_2', c_3',\ldots, c_N',\}$, there are $r$ different permutations, $r= A_T^N =T(T-1)(T-2)\ldots(T-N+1)$. Each permutation corresponds to a sequence $s_i$, $s_i \in \{0,1\}^l$, $l=\log_2 r$. Therefore, there is a mapping relation: $\{c_1', c_2', c_3',\ldots, c_N',\} \to s_i$, $s_i$ constitutes a set of stego-sequence: $S=\{s_1, s_2, s_3,\ldots, s_r\}$.

3. Secret message collection $M$: $M=\{m_1, m_2, m_3,\ldots, m_r\}$, $m_i \in \{0,1\}^l$. The key $k \in \mathcal{K}$, $\mathcal{K}=\{0,1\}^l$, $|\mathcal{K}|=2^l$;

Data hiding:

$F_{stego}(m_i, k)$ is equivalent to a map process: $M \to S \to X$,

1) $s_j = m_i \oplus k$.

2) $s_j \xrightarrow{Mapping} \{c_1', c_2', c_3',\ldots, c_N',\}$.

Then send the image sequence $c'=\{c_1', c_2', c_3',\ldots, c_N',\}$ to the receiver.

Data extraction:

$F_{extract}(c_j, k)$ is equivalent to another map process: $X \to S \to M$,

1) $\{c_1', c_2', c_3',\ldots, c_N',\} \xrightarrow{Mapping} s_j$.

2) $m_i = s_j \oplus k$.

*b. Security*

The original cover in coverless steganography should be the library $X$. Any image from $X$ is natural and will not expose the existence of steganography. The message is not embedded into an image. Instead, it is represented by the changes of the natural image sequence. Namely, the entire natural image library $X$, rather than a certain image, is the original cover. And the sequence of images from *X* corresponding to a particular arrangement is a stego-cover.

Under SCOA, the attacker can obtain some images from $\{c_1', c_2', c_3',\ldots, c_N',\}$, and the individual image $c_i'$ is natural. It is impossible to statistically analyze these images to distinguish the unnatural cover. Therefore, for the attacker, that $D(P(c), P(c'))=0$ is satisfied. The message was XOR encrypted in Step 1 of the data hiding process by using the key *k*. Without *k*, $m_i$ cannot be obtained in Step 2 of the extraction process. The key space $|\mathbb{K}|$ is $2^l$, and $|\mathbb{K}|$ can meet certain computational complexity requirement when *l* is large enough. Therefore, coverless steganography is stego-secure against SCOA.

Under KCA, the finite images in *X* are exposed to the attacker. In the learning phase, the attacker could find that the exchanging image $c_i'$ comes from *X*. In challenge phase, the attacker might directly suspect the communication if the exchanging images are still from *X*. Therefore, coverless steganography does not resist KCA.

*c. Practical requirements*

According to the security analysis, the security requirements of the coverless steganography instance in practice are as follows.

1. The parts to be kept secrecy include: the image library *X*, and the key *k*.

2. $|\mathbb{K}|$ meets the certain requirement of computational complexity.

3. The parameters *N*, *T* and *x* which denotes the number of times that the algorithm can be securely used or the number of the available stego-sequences should satisfy the following security requirement:

Assuming a *case* that the attacker obtains all the *T* images in *X* through *x* stego-sequences randomly, the probability of that case is record as $\Pr_X(x, N, T)$, $(T< xN)$. To obtain $\Pr_X$, we first calculate the probability $\Pr_{\overline{X}}$, that there are still *y* images remaining secret to the attacker：

$$\Pr_{\overline{X}}(y,x,N,T) = C_T^y \cdot \left(\frac{C_{T-y}^N}{C_T^N}\right)^x \quad (12)$$

Then $\Pr_X$ can be obtained,
$\Pr_X(x, N, T) =$

$$1-\sum_{y=1}^{T-N}\Pr_{\overline{X}}(y,x,N,T)^x = 1-\sum_{y=1}^{T-N} C_T^y \cdot \left(\frac{C_{T-y}^N}{C_T^N}\right)^x \quad (13)$$

When $N \ll T$, it is difficult for an attacker to forecast all images in *X* through several stego-sequences. In practice, the security requirement on *N*, *T*, and *x* is to ensure $\Pr_X(x, N, T)$ small enough, i.e., $\Pr_X(x, N, T) < \zeta$, $\zeta$ is an arbitrary small value.

2) GAN BASED GENERATIVE STEGANOGRAPHY

The main function of generative adversarial networks under the semi-supervised state is to generate a controllable result that fully follows with the distribution characteristics of the training dataset. It has natural advantage to use GAN for designing steganographic methods, by which the generated stego-cover follows the distribution of the training covers. This paper believes that there are two constructing frameworks: one is cover first generative steganography (CFGS) to generate the secret message based on a fixed cover [34], and the other is message first generative steganography (MFGS) to generate the stego-cover based on the fixed desired message.

*a Cover first generative steganography*

In CFGS, a natural image is arbitrarily selected as the cover, and the image can be used to generate a secret message under the control of the key without any modification. The receiver can generate the secret message if and only if the cover image and the key are obtained at the same time. In [34], two GANs are used to implement CFGS: Message-GAN with two networks $G_M$ and $D_M$, and Cover-GAN with three networks $G_C$, $R_C$ and $A_C$.

$G_M$ in Message-GAN is used by the receiver to generate message *m* by inputting the message feature code *f*. For Cover-GAN, the sender uses $G_C$ to generate the key *k* by inputting the cover *C* while the receiver uses $R_C$ to obtain *f*, then the message can be obtained through $G_M$.

(1) Algorithm

Data hiding:

1. Randomly select a natural image as the cover, *C*.

2. Feature extraction: *f* =*Feature_Extraction*(*m*).

3. Random 01 sequence is denoted as *r* with length *l*, *l*=*length*(*r*) = *length* (*f*), and then calculate *f'*= *r* ⊕*f*.

4. $G_C(C, f') = r_c$, and the key is obtained: $k(r, r_c)$.

Data extraction:

1. Obtain the feature code *f* by using *C* and *k*:

$$R_C(C, r_c) = f' \quad (14)$$
$$f = r \oplus f' \quad (15)$$

2. Generate the message by using $G_M$:

$$m = G_M(f) \quad (16)$$

(2) Security

In CFGS, the function of the cover image is to conceal the existence of the hidden message and to act as the seed for driving the generators. Therefore, it satisfies that $D(P(C), P(C'))=0$ both in SCOA and KCA. Since the feature code *f* is XOR encrypted by *k*, the message cannot be obtained without *k* due to the Eqs. (14-16). The key space $|\mathbb{K}|=2^l$, and $|\mathbb{K}|$ can meet certain computational complexity requirement when *l* is large enough. Therefore, coverless steganography is stego-secure against SCOA and KCA.

However, $r_c$ of *k* comes from the data hiding process, thus resulting in its limited applications. The communication requires the transfer of *k*. It is somehow similar to cryptography, in which it usually introduces another data expansion (ciphertext expansion and key expansion) to transfer some messages by exchanging the two random-like data. Their difference lies in the undetectable part (the natural cover) in the CFGS while ciphertext or the key in cryptosystem are both in the form of random noise, sensitive for attackers. From this point of view, CFGS combines some technical features of cryptography and steganography.

(3) Practical requirements

The security requirements of CFGS in practice are as following:

1. The part to be kept secrecy is the key $k$.

2. $|\mathbb{K}|$ meets the certain requirement of computational complexity.

3. The key $k$ and the cover $C$ cannot be obtained by the attacker at the same time. $C$ can be publicly transmitted, because it is not modified. Here, $k$ is not required to be transmitted through the secret channel, because its content is in the random state and does not reveal any information about the message content.

*b. Message first generative steganography*

MFGS aims to generate a stego-cover containing the message by a generator. This generator can be obtained based on GAN. The distribution characteristics of the generated cover can be constrained by the GAN to be indistinguishable. In this paper, we propose an instance of MFGS.

In the instance, the starting point of constructing the cover is the fixed message $m$. The goal is to generate a natural stego-cover $C'$ based on the inpainting GAN in [35], in which a corrupted image is inputted for inpainting according to the peripheral content of the residual image. Several positions from the corrupted part or the residual part are fixed in advance for carrying secret messages. In the inpainting process, not only the distribution characteristics of the natural image need to be guaranteed, but also the message at the embedding position cannot be destroyed.

$$C' = Gen(m) \qquad (17)$$

where $Gen(.)$ represents an image generation method and $C'$ satisfied:

Requirement 1: $m = F_{extract}(C')$ (18)

Requirement 2: $D(P(C), P(C'))=0$ (19)

Requirement 1 guarantees the accuracy of the message extraction. Requirement 2 ensures the same distribution that the cover and the stego-cover follow. The generated $C'$ is not any an existing image from the training set, but it can theoretically and practically guarantee its distribution follows the distribution of the training images.

(1) Algorithm

Data hiding:

1. Randomly select a natural image $I$ of size $h \times h$ and a mapping matrix $M \in \{0,1\}^{h \times h}$ of size $h \times h$. $M$ is used to mark the embedded positions. If the value is "0", it indicates that the pixel at the corresponding position of $I$ has no information embedded. If the value is "1", the pixel carries additional information. Here, to drive the work of complementing GAN, we randomly corrupt the image $I$, and obtain the corrupted image $I_c$.

2. The secret message $m$ is XOR-encrypted using a random sequence $r$. The length of $m$ and $r$ is denoted as $v$:

$$s = m \oplus r \qquad (20)$$

3. Replace the LSB (least significant bit) of the pixel at embedded position in $I_c$, and obtain an embedded corrupted image $I_c'$:

$$I_c' = Replace(LSB(I_c' \odot M), s) \qquad (21)$$

Replace (.) represents a bit-wise replacement; $. \odot M$ represents to locate the pixels at the embedded positions according to the mapping matrix $M$.

4. Input $I_c'$ into the inpainting GAN. Finally, the complement image $C'$ is obtained. The LSB of the pixel at the embedded position in $C'$ is required to be fixed during inpainting. $C'$ is the stego-cover and the key $k$ is ($r$, $M$).

Data extraction:

With $C'$ and $k$ ($r$, $M$), the receiver uses the mapping matrix $M$ to extract the LSB of the pixel at embedded position in $C'$ and finally obtains the message $m'$.

$$s' = LSB(C' \odot M) \qquad (22)$$
$$m' = s' \oplus r \qquad (23)$$

The working flow of MFGS is shown in Fig. 1.

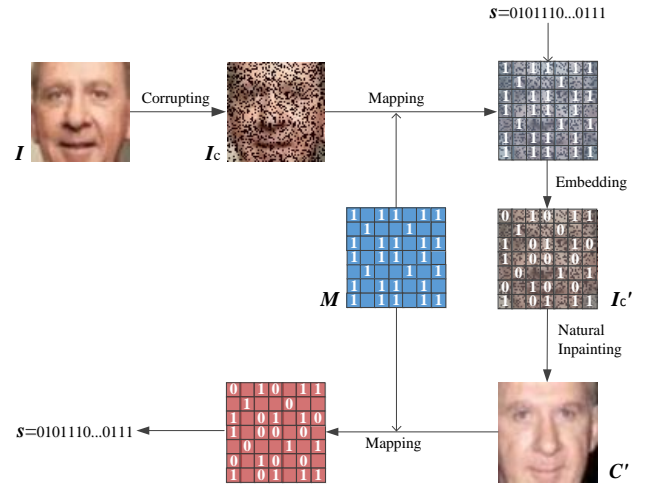

**FIGURE. 1. Working flow of MFGS.**

(2) Training architecture

The generator is the crucial piece of MFGS. The training goal is to meet the following two conditions in Eqs. (19)(24).

$$LSB(C' \odot M) = m \qquad (24)$$

GAN generates artificial samples that are indiscernible from the real counterparts via the competition between a generator ($G$) and a discriminator ($D$), i.e., alternating the maximization and minimization steps in Eq. (25).

$$\min_G \max_D V(D,G) = E_{C \sim P_{data}}\left[\log D(C)\right] + E_{x \sim P_{noise}}\left[\log(1-D(G(x)))\right] \qquad (25)$$

where $D(C)$ is the probability that $C$ is a real image from the training dataset rather than synthetic, and $G(x)$ is a synthetic image for input $x$. GAN will finally reach a state of Nash equilibrium of $G$ and $D$. The performances of $G$ and $D$ both get promoted and there is a 50% chance for $D$ to distinguish real samples from generated ones by $G$. Next, we analyze how to meet the two conditions.

Condition 1 (in Eq. (19)), we find the first and second derivative of Eq. (25) to obtain the theoretical value of the optimal value by the discriminator $D$ [36]:

$$D^*_G = \frac{P_{\text{data}}(x)}{P_{\text{data}}(x) + P_G(x)} \quad (26)$$

Eq.(26) is the theoretical best value that the discriminator can achieve. For the generator, the best theoretical result is $P_{\text{data}}(x) = P_G(x)$, i.e., $D(P(C), P(C')) = 0$, $D^*_G = 1/2$. Then $V = -\log 4$.

During the training, we use Deep Convolutional GAN (DCGAN) [37] for optimization. For $G$ and $D$, we fix one to obtain the best parameter setting of the other iteratively:

$$D_G = \arg\max_G V(G, D) \quad (27)$$

$$G_D = \arg\max_D V(G, D) \quad (28)$$

In practical training, $G$ is based on the back-propagation of $V$ to optimize the parameters. When the training iterations are enough, the extreme value of $G$ under the fixed $D$ can always be obtained. If $D^*_G$ is substituted into $V(D, G)$, it is possible to calculate the extreme value of $V$ on $G$, $V^*(D^*_G, G)$:

$$V(D^*_G, G) =$$

$$E_{C \sim P_{\text{data}}}\left[\log \frac{P_{\text{data}}(C)}{P_{\text{data}}(C) + P_G(C)}\right] + E_{x \sim P_{\text{noise}}}\left[\log\left(1 - \frac{P_{\text{data}}(x)}{P_{\text{data}}(x) + P_G(x)}\right)\right] =$$

$$-\log 4 + 2 D_{JSD}(P_{\text{data}}, P_G) \geq -\log 4 \quad (29)$$

$V^*(D^*_G, G) \geq -\log 4$. And iff $D_{JSD}(P_{\text{data}}, P_G) = 0$, $V^*(D^*_G, G) = -\log 4$. In our experiments, DCGAN's final training will reach the extreme value of $V$. At that time, its JSD distance is 0. We can believe that the distributions has been fitted, because if $D_{JSD}(P_{\text{data}}, P_G) = 0$, and we substitute it into Eq.(29), $V^*(D^*_G, G) = -\log 4$, which is the same as the theoretical value of $V$ when $D(P(C), P(C')) = 0$.

Condition 2 (in Eq. (24)), the generator parameters trained by DCGAN are no longer changed. The noise $z$ is inputted into $G$ to generate an image. To satisfy condition 2, we introduce the L1 distance to measure the extracted message distortion.

$d$ on the input $(X, X')$ is introduced to show the distortion between two sequences $X$ and $X'$

$$d(X, X') = \sum_{i=1}^{N} |x_i - x_i'| \quad (30)$$

This is not to measure the distribution of the two samples, but to specifically measure the bit accuracy of the extracted messages, so L1 distance is used to establish the message distortion, $L_{\text{message}}$:

$$L_{\text{message}} = d(s, s') \quad (31)$$

Then we modify the input $z$ to optimize $L_{\text{message}}$ [38], when iteratively updating $z$ using back-propagation by

$$z \leftarrow z - \gamma_z \nabla_z L_{\text{message}}, \quad \nabla_z L_{\text{message}} = -\frac{\partial L_{\text{message}}}{\partial z} \quad (32)$$

Through enough training, the distortion $L_{\text{message}}$ theoretically reaches a minimum value of 0. The MFGS generator is obtained.

(3) Experimental Results

We implement our adversarial training on LFW datasets which contains more than 13,000 images of human faces. 12,000 sample images are used for the training while approximately other 1000 sample images are removed from the dataset for testing on a NVIDIA Titan XP GPU card. The generative network $G$ takes a random 100 dimensional vector from uniform distribution among [-1,1] to output a $64 \times 64 \times 3$ image. The discriminator model $D$ is structured essentially in reverse order. The input layer is an image with $64 \times 64 \times 3$ dimensions, and the output layer has a two-class feedback to minish the loss. As shown in Fig. 2, the original images, the corrupted images, stego corrupted images, and generated stego images by 500 and 1000 iterations of updating $z$ are respectively at column 1-5. The generated stego images in Fig. 2(a) have a high perceptual quality. With the corruption degree on the context growing over 50%, the perceptual quality of the generated stego images tends to decline.

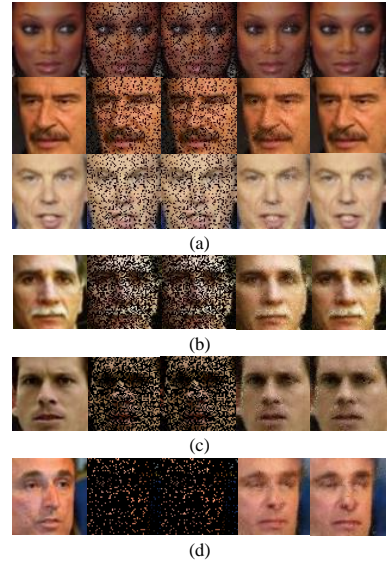

(a)

(b)

(c)

(d)

**FIGURE 2.** Instance of MFGS processes: original images, corrupted images, stego corrupted images, and generated stego images after 500 iterations and 1000 iterations. (a) 0.8bpp capacity and 20% missing of context; (b) 0.5bpp capacity and 50% missing of context; (a) 0.2bpp capacity and 80% missing of context; (a) 0.1bpp capacity and 90% missing of context.

We analyze MFGS using the blind steganalyzer in spatial domain and the ensemble classifier. 686-dimensional SPAM features [39] and 5404-dimensionnal SCRMQ1 features [40] with ensemble classifiers [41] are implemented as random forests in this experiment. As Fig. 3 shows, the testing errors arise with increasing training iterations. All the 1000 stego-covers and 1000 original covers are generated at 1000 iterations from corrupted images by the image inpainting. The images are divided randomly into two halves, one used for training and the other for testing. In Fig. 3, we show the testing errors of the payloads from 0.1bpp to 0.5bpp (bits per pixel). The testing errors of MFGS are higher than HOGO or HILL when the payloads are below 0.2bpp. The errors descend rapidly with a high payload due to the hardness of the convergence of GAN to eliminate $L_{message}$. And the message extraction accuracy changes with the embedding bit planes, as shown in Fig. 4.

MFGS has the advantage of security guarantee in theory, however, the practical performance of MFGS is still inferior to state-of-the-art steganographic algorithms, which might be improved by introducing high-performance GANs.

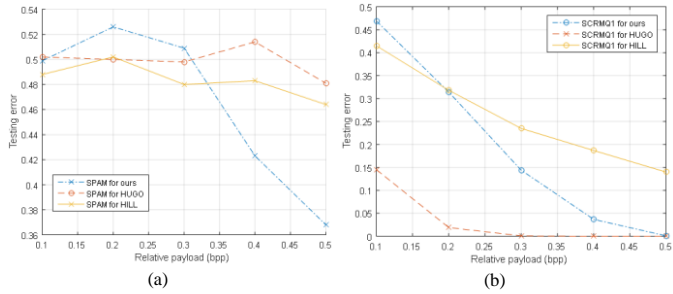

**FIGURE 3.** Error rates for five different payloads (0.1-0.5bpp) with HUGO, HILL and our method using (a) SPAM features and (b) SCRMQ1 features.

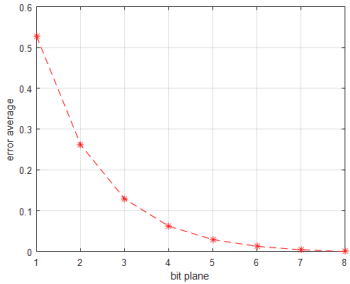

**FIGURE 4.** Average extraction error rates for different embedding bit planes.

(4) Security

According to the inpainting principle and experimental results of GAN, the generated image $C'$ is not the image $I$, but another new image that fits the distribution characteristics of the training image dataset. The role of $I$ is to drive the natural completion. The distribution characteristics of the generated images come from the training dataset, and rely on GAN's powerful learning ability. Therefore, the original cover in MFGS is not a single image, but a type of image group with certain distribution characteristics. Such characteristics might be quite complex and difficult to describe by establishing a model. There should be an infinite number of theoretical images belonging to the group, but they cannot represent all natural images.

Without knowing training dataset, it is difficult to describe the characteristics or infer the training dataset backwards through a number of generated images or a trained GAN. In the absence of the key, the message cannot be extracted due to Eqs. (17-18). Key space $|\mathbb{K}|=2^{hh} \times 2^v = 2^{hhv}$, and $|\mathbb{K}|$ meets the certain computational complexity when $v$ and $h$ are large enough. Therefore, MFGS is stego-secure against SCOA.

However, the number of the training images and the training iterations is limited. Therefore, the characteristics that can be reflected are also limited. In KCA, the training set should be exposed to the attacker. In the learning phase, the attacker can use the GAN tool to train a discriminator capable of distinguishing the image characteristics through a purposefully expanded training set. In challenge phase, if the images exchanged in the channel always conform to the characteristics of the known dataset, it can be considered suspectable. Therefore, MFGS is not stego-secure against KCA.

(5) Practical requirements

The security requirements of MFGS in practice are as follows.

1. The parts to be kept secrecy include: the training dataset, and the key $k$.

2. $|\mathbb{K}|$ meets the certain requirement of computational complexity.

### B. Stego-Security against KCA

#### 1) ALGORITHM

The instance of the stego-secure algorithm against KCA is the LWE-based RDH-ED [42]. Ciphertext has the advantage to act as the cover because of its simple model, the randomly uniform distribution. The security focus would be to ensure the stego-ciphertext evenly distributed. Of course, the practical significance of utilizing methods of RDH-ED for steganography remains controversial currently. Here, we assume that ciphertext can act as a steganographic cover, mainly from the theoretical perspective to analyze the security.

In [42], controllable redundancy of LWE cryptosystem is analyzed and utilized for data hiding. The encryption blowup of LWE comes from the quantization vector in decryption, which is not controllable for attackers while the LWE encryption key keeper can control it. The range of a quantization element in the quantization vector takes up half of the integer domain $\mathbb{Z}_q$ ($q$ is a big Integer) to recover 1 bit plaintext, i.e., a quantization element with $q/2$ possible values only corresponds to 1 bit of plaintext. Scheme [42] takes advantage of the redundancy of quantization element to embed data. There are more detail descriptions in [42] about the methodology and theoretical analysis. In this section, we concentrate on its security in KCA.

#### 2) SECURITY

In KCA, the attacker has access to severer pairs of the original cover $C$ and its corresponding stego-cover $C'$. As for the LWE based RDH-ED, the original cover is LWE encrypted data which follows a random uniform distribution, a straightforward and unified distribution. A crucial criterion of the security in [42] is to ensure that the marked ciphertext (stego-cover) should follow the same distribution to maintain the hardness of LWE encryption and the undetectability. The probability distribution function (PDF) and statistical features after embedding, including histogram, mean, and the average information entropy, have been analyzed to prove that. Therefore, in KCA, LWE based RDH-ED meets $D(P(C), P(C'))=0$. According to [42], key space $|\mathbb{K}|$ could meet the security with resistance to quantum computing analysis. Therefore, the example in [42] has reached the stego-security against KCA.

In CCA, an attacker can invoke the current key several times to perform an embedding or extraction operation. Therefore, all the possible open parameters at each step of these operations should not reveal any clues to the existence of steganography. However, the example in [42] cannot ensure that, because the quantization vectors from original ciphertext and stego-ciphertext would follow different distributions (shown in Fig. 5, there are more peaks in Fig. 5c-5d than in Fig. 5a-5b), which is decided by the data hiding levels and the sampling of keys. In RDH-ED, there are two types of keys: encryption/ decryption key and data-hiding key. The decryption key should be secrecy in the cryptosystem, but when analyzing the stego-security from the perspective of the steganography system, we can assume that the encryption key can be assumed to be compromised by the steganalysis attacker, and only the data- hiding key is secret. In conclusion, the threat in CCA lies in that the decryption process could expose the presence of data hiding.

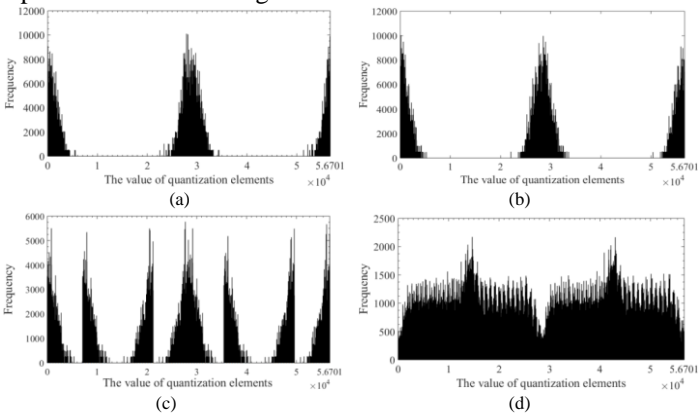

**FIGURE 5.** Distributions of quantization vectors from original ciphertext and stego-ciphertext with q=56701: (a)-(b) Original ciphertext; (c) Stego-ciphertext after once data hiding; (d) Stego-ciphertext after multilevel data hiding.

In the CCA learning phase, the attacker can learn the differences of distribution characteristics of the quantized vectors in the decryption or message extraction stage. In the challenge phase, it obtains the quantization vectors by using only part of the decryption key, and then it could confirm the presence of the steganography with a large correct rate. Therefore, the example in [42] cannot reach the stego-security against CCA.

3) PRACTICAL REQUIREMENTS

1. The parts to be kept secrecy include: the decryption key and the data-hiding key. Specifically speaking, only the secret key of LWE owner should operate data hiding because of his access to the controllable redundancy. To improve the security, the decryption key could be transmitted to others by using some key distribution strategy or introducing LWE based proxy re-encryption technology [42];

2. $|\mathbb{K}|$ meets the certain requirement of computational complexity.

### C. Stego-Security against CCA

The stego-security against CCA requires that the steganography system has no possibility of expose the presence of steganography in the various operations. That is, the process parameters that may be disclosed to the attacker do not provide any content that can be differentiated or directly related to the hidden data itself. With regard to the instance against CCA, it can be a RDH-ED based steganography that supports third-party embedding. The original ciphertext and the stego-ciphertext are both random-evenly distributed. The data hider and the plaintext content holder can be different persons. Each of them is assigned a key independently. The processes of message extraction and the plaintext decryption are completely separable.

The RDH-ED based on the self-blinding Paillieris algorithm in [43] can be an instance here if it is used to implement steganography. It needs to calculate a self-blinding parameter first, and then use different keys and quantization methods to perform the decryption or the extraction operation. Paillier algorithm is used for encrypting. After embedding, the ciphertext maintains a good random distribution in the absence of the data-hiding key, even if the decryption key is available. In CCA, the self-blinding parameters are consistent with random and will not expose the presence of steganography. Only when the data-hiding keys are mastered, can the parameters indicate the secret messages. Therefore, this method can achieve the stego-security against CCA.

However, the Paillier encryption algorithm itself cannot resist the *adaptive chosen ciphertext attack* (ACCA, or CCA2) [44], and the ciphertext is the cover in the stego-system. Therefore, in ACCA of steganalysis, the Paillier encryption based algorithm cannot guarantee plaintext security, and will expose the existence of the hidden data in the plaintext with a large possibility.

### D. Stego-Security against ACCA

Currently, the instance against ACCA has not yet an available method. It needs the data hiding method to resist the attacker's arbitrary invoking of the system. Namely, during the repeated attack attempts, no attack experience will be accumulated. To achieve this kind of security, this paper believes that there might be (not limited to) several possible methods:

1. To introduce RDH-ED based on cryptography against CCA2 to implement steganography.

2. Steganography based on quantum information hiding;

3. To set the number and type of training images, and the training iterations big enough in MFGS to meet certain security requirements.

## VII. CONCLUSION

This paper discusses several theoretical issues and principles in steganography security. We complement the setup of the attacker, Eve, in the steganography model that Eve could be given more active privileges only to complete a passive attacking goal. Therefore, stego-security requires the system should ensure (passive or active) Eve cannot complete the two attack goals in Definition 1. The indistinguishability of cover and stego-cover is the first guarantee of stego-security. In practice then, stego-security should be key-secrecy based. We formalize the necessary conditions of stego-security according to Kerckhoffs' principle in steganography. Finally, algorithm instances of the four security levels are presented and analyzed. This paper focuses on theoretical analysis of security, aims to provide theoretical reference for the future developments or improvements of the steganography technology.

Furthermore, there are another two views of the future research on stego-security:

1. To improve keys' distributing by further introducing the mechanism of public key system into steganography.

2. To achieve the check of the integrity of the secret message, as well as the authentication of the source of the hidden message.

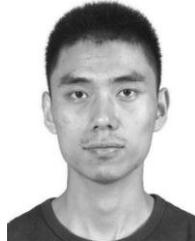

**Yan Ke** received the B.S. degree in information research & security from Engineering University of Chinese People Armed Police Force (PAP), Xi'an, China in 2014 and the M.S. degree in cryptography from Engineering University of PAP, Xi'an, China in 2016. He is currently pursuing the Ph.D. degree in cryptography at Engineering University of PAP. His research interest includes information hiding, lattice cryptography, deep learning.

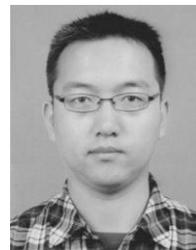

**Jia Liu** received the M.S. degree in cryptography from Engineering University of PAP, Xi'an, China, in 2007 and Ph.D. degree in neural network and machine learning from Shanghai Jiao Tong University, Shanghai, China, in 2012. Currently, he has been with the Key Laboratory of Network and Information Security under PAP as an associate professor. His research interests include pattern recognition and image processing.

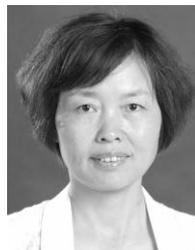

**Minqing Zhang** received the M.S. degree in computer science & application from Northwestern Polytechnical University, Xi'an, China, in 2001, and Ph.D. degree in network & information security from Northwestern Polytechnical University, Xi'an, China, in 2016. Currently, she has been with the Key Laboratory of Network and Information Security under Chinese People Armed Police Force as a professor. Her research interests include cryptography, and trusted computation.

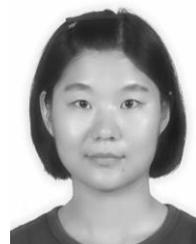

**Tingting Su** received the B.S. degree in information research & security from Engineering University of PAP, Xi'an, China in 2010 and the M.S. degree in cryptography from Engineering University of PAP, Xi'an, China in 2013. Her research interest includes mathematics statistics and cryptography.

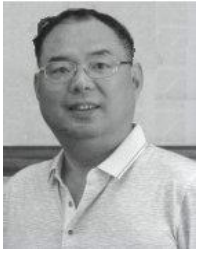

**Xiaoyuan Yang** received the B.S. degree in applied mathematics from Xidian University, Xi'an, China, in 1982 and the M.S. degree in cryptography & encoding theory from Xidian University, Xi'an, China, in 1991. Currently, he has been with the Key Laboratory of Network and Information Security under PAP as a professor. His research interests include cryptography and trusted computation.